\magnification=\magstep1 
\font\bigbfont=cmbx10 scaled\magstep1
\font\bigifont=cmti10 scaled\magstep1
\font\bigrfont=cmr10 scaled\magstep1
\vsize = 23.5 truecm
\hsize = 15.5 truecm
\hoffset = .2truein
\baselineskip = 14 truept
\overfullrule = 0pt
\parskip = 3 truept
\def\frac#1#2{{#1\over#2}}

\nopagenumbers
\topinsert
\vskip 3.2 truecm
\endinsert
\centerline{\bigbfont BEC, BCS AND BCS-BOSE CROSSOVER THEORIES} 
\vskip 6 truept
\centerline{\bigbfont IN SUPERCONDUCTORS AND SUPERFLUIDS}
\vskip 20 truept
\centerline{\bigifont M. de Llano,$^{a}$ F.J. Sevilla,$^{b,c}$ M.A. Sol\'{\i}s$^{b}$ \&
J.J. Valencia$^{a,d}$}
\vskip 8 truept
\centerline{\bigrfont$^{a}$Instituto de Investigaciones en Materiales,
UNAM, 04510 M\'{e}xico, DF, Mexico} 
\vskip 2 truept
\centerline{\bigrfont $^{b}$Instituto de F\'{\i}sica, UNAM, 01000 M\'{e}xico, DF, 04510 Mexico}
\vskip 2 truept
\centerline{\bigrfont $^{c}$Consortium of the
Americas for Interdisiplinary Science,}
\vskip 2 truept
\centerline{\bigrfont University of New Mexico 
Albuquerque, NM 87131, USA}
\vskip 2 truept
\centerline{\bigrfont $^{d}$Universidad de la Ciudad de M\'exico,}
\vskip 2 truept
\centerline{\bigrfont  San Lorenzo Tezonco, 09940 M\'{e}xico, DF, Mexico}

\vskip 1.8 truecm

\centerline{\bf 1.  INTRODUCTION}
\vskip 12 truept

Though commonly unrecognized, a superconducting BCS condensate consists of
equal numbers of two-electron (2e) and two-hole (2h) Cooper pairs (CPs).\ A 
{\it complete} {\it boson-fermion} (statistical) {\it model}
(CBFM), however, is able to depart from this perfect 2e-/2h-CP symmetry and
yields [1] robustly higher $T_{c}$'s without abandoning
electron-phonon dynamics mimicked by the BCS/Cooper model interaction $V_{%
\bf{k,k^{\prime }}}$ which is a nonzero negative constant $-V,$ if and
only if single-particle energies $\epsilon _{k},\epsilon _{k^{\prime }}$ are
within an interval $[$max$\{0,\mu -\hbar \omega _{D}\},\mu +\hbar \omega
_{D}]$ where $\mu $ is the electron chemical potential and $\omega _{D}$ is
the Debye frequency. The CBFM is ``complete'' only in the sense that 2h-CPs
are not ignored, and reduces to all the known statistical theories of
superconductors (SCs), including the BCS-Bose ``crossover'' picture but goes
considerably beyond it.

Boson-fermion (BF) models of SCs as a Bose-Einstein condensation (BEC) go
back to the mid-1950's [2-5], pre-dating even the
BCS-Bogoliubov theory [6-8]. Although BCS theory only
contemplates the presence of ``Cooper correlations'' of single-particle
states, BF models [2-5, 9-17] posit the
existence of actual bosonic CPs. Indeed, CPs appear to be universally
accepted as the single most important ingredient of SCs, whether
conventional or ``exotic'' and whether of low- or
high-transition-temperatures $T_{c}$. In spite of their centrality, however,
they are poorly understood. The fundamental drawback of early [2-5] 
BF models, which took 2e bosons as analogous to diatomic
molecules in a classical atom-molecule gas mixture, is the notorious absence
of an electron energy gap $\Delta (T)$. ``Gapless'' models cannot describe
the superconducting state at all, although they are useful [16,17]
 in locating transition temperatures if approached from above, i.e., $%
T>T_{c}$. Even so, we are not aware of any calculations with the early BF
models attempting to reproduce any empirical $T_{c}$ values. The gap first
began to appear in later BF models [9-14]. With two [12, 13] 
exceptions, however, all BF models neglect the effect of 
{\it hole} CPs accounted for on an equal footing with electron CPs,
except the CBFM which consists of {\it both} bosonic CP species
coexisting with unpaired electrons, in a {\it ternary} gas mixture.
Unfortunately, no experiment has yet been performed, to our knowledge, that
 distinguishes between electron and hole CPs.

The ``ordinary'' CP problem [18] for two distinct interfermion
interactions (the $\delta $-well [19, 20] or the
Cooper/ BCS model [6, 18] interactions) neglects the effect of
2h CPs treated on an equal footing with 2e\ [or, in general, two-particle
(2p)] CPs. On the other hand, Green's functions [21] can naturally
deal with hole propagation and thus treat both 2e- and 2h-CPs [22, 23].
 In addition to the generalized CP problem, a crucial result %
[12, 13] is that the BCS condensate consists of {\it equal
numbers of 2p and 2h CPs}. This was already evident, though widely ignored,
from the perfect symmetry\ about $\epsilon =\mu $ of the well-known
Bogoliubov [24] $v^{2}(\epsilon )$\ and $u^{2}(\epsilon )$
coefficients, where $\epsilon $ is the electron energy.

Here we show: a) how the crossover picture $T_{c}$s, defined
self-consistently by {\it both} the gap and fermion-number equations,
requires unphysically large couplings (at least for the Cooper/BCS model
interaction in SCs)\ to differ significantly from the $T_{c}$ from ordinary
BCS theory defined {\it without }the number equation since here the
chemical potential is assumed equal to the Fermi energy; how although
ignoring either 2h- {\it or }2e-CPs in the CBFM b) one obtains the
precise BCS gap equation for all temperatures $T$, but c) only {\it half }%
the $T=0$\ BCS condensation energy emerges. The gap equation gives $\Delta
(T)$ as a function of coupling, from which $T_{c}$ is found as the solution
of $\Delta (T_{c})=0$.\ The condensation energy is simply related to the
ground-state energy of the many-fermion system, which in the case of BCS is
a rigorous upper bound to the exact many-body value for the given
Hamiltonian.\ Results (b) and (c) are also expected to hold for
neutral-fermion superfluids (SFs)---such as liquid $^{3}$He [25, 26], 
neutron matter and trapped ultra-cold fermion atomic gases %
[27-38]---where the pair-forming two-fermion
interaction of course differs from the Cooper/BCS one for SCs.

\vskip 28 truept

\centerline{\bf 2.  THE COMPLETE BOSON-FERMION MODEL}
\vskip 12 truept

The CBFM [12, 13] is described in $d$ dimensions by the Hamiltonian $%
H=H_{0}+H_{int}$. The unperturbed Hamiltonian $H_{0}$\ corresponds to a
non-Fermi-liquid ``normal'' state, being an {\it ideal }(i.e.,
noninteracting) ternary gas mixture of unpaired fermions and both types of
CPs namely, 2e and 2h. It is%
$$
H_{0}=\sum\limits_{\bf{k}_{1},s_{1}}\epsilon _{\bf{k}_{\bf{1}%
}}a_{\bf{k}_{1},s_{_{1}}}^{+}a_{\bf{k}_{1},s_{_{1}}}+\sum\limits_{%
\bf{K}}E_{+}(K)b_{\bf{K}}^{+}b_{\bf{K}}-\sum\limits_{\bf{K}%
}E_{-}(K)c_{\bf{K}}^{+}c_{\bf{K}}
$$
where as before $\bf{K\equiv k}_{\bf{1}}+\bf{k}_{\bf{2}}$ is
the CP center-of-mass momentum (CMM) wavevector while $\epsilon _{\bf{k}%
_{1}}\equiv \hbar ^{2}k_{1}^{2}/2m$ are the single-electron, and $E_{\pm }(K)
$\ the 2e-/2h-CP {\it phenomenological, }energies.\ Here $a_{\bf{k}%
_{1},s_{1}}^{+}$ ($a_{\bf{k}_{1},s_{1}}$) are creation (annihilation)
operators for fermions and similarly $b_{\bf{K}}^{+}$ ($b_{\bf{K}}$)
and $c_{\bf{K}}^{+}$ ($c_{\bf{K}}$) for 2e- and 2h-CP bosons,
respectively. Two-hole CPs are considered {\it distinct }and{\it \
kinematically independent }from 2e-CPs.

The interaction Hamiltonian $H_{int}$ (simplified by dropping all $\bf{K}%
\neq 0$ terms, as is done in BCS theory in the {\it full }Hamiltonian but
kept in the CBFM in $H_{0}$) consists of four distinct BF interaction
vertices each with two-fermion/one-boson creation and/or annihilation
operators. The vertices depict how unpaired electrons (subindex +) [or holes
(subindex $-$)] combine to form the 2e- (and 2h-) CPs assumed in the $d$%
-dimensional system of size $L$, namely 
$$
H_{int}=L^{-d/2}\sum\limits_{\bf{k}}f_{+}(k)\{a_{\bf{k},\uparrow
}^{+}a_{-\bf{k},\downarrow }^{+}b_{\bf{0}}+a_{-\bf{k},\downarrow
}a_{\bf{k},\uparrow }b_{\bf{0}}^{+}\}
$$
$$
+L^{-d/2}\sum\limits_{\bf{k}}f_{-}(k)\{a_{\bf{k},\uparrow }^{+}a_{-%
\bf{k},\downarrow }^{+}c_{\bf{0}}^{+}+a_{-\bf{k},\downarrow }a_{%
\bf{k},\uparrow }c_{\bf{0}}\}  \eqno{(1)}
$$%
where $\bf{k}\equiv \frac{1}{2}\bf{(k}_{\bf{1}}-\bf{k}_{%
\bf{2}})$ is the relative wavevector of a CP. The interaction vertex
form factors $f_{\pm }(k)$ in (1) are\ essentially the Fourier
transforms of the 2e- and 2h-CP intrinsic wavefunctions, respectively, in
the relative coordinate of the two fermions. In Refs. [12, 13]
they are taken as
$$
f_{\pm}(\epsilon )=\left\{ 
\eqalign{
f & \quad {\rm if} \quad
\frac12
[E_{\pm}(0)-\delta \varepsilon ] < \epsilon <%
{\frac12}
[E_{\pm }(0)+\delta \varepsilon ] \cr 
0 & \quad {\rm otherwise.}%
}
\right. \eqno{(2)}
$$%
One then introduces the quantities $E_{f}$ and $\delta \varepsilon $ as 
{\it new} phenomenological dynamical energy parameters (in addition to
the positive BF vertex coupling parameter $f$) that replace the previous $%
E_{\pm }(0)$ parameters, through the definitions 
$$
E_{f}\equiv \frac{1}{4}[E_{+}(0)+E_{-}(0)] \quad {\rm and} \quad \delta
\varepsilon \,\equiv \,\frac{1}{2}[E_{+}(0)-E_{-}(0)]  \eqno{(3)}
$$%
where $E_{\pm }(0)$ are the (empirically {\it un}known) zero-CMM energies
of the 2e- and 2h-CPs, respectively. Alternately, one has the two relations 
$$
E_{\pm }(0) = 2E_{f}\pm \delta \varepsilon . 
 \eqno{(4)}
$$
The quantity\ $E_{f}$ serves as a convenient energy scale; it is not to be
confused with the Fermi energy $E_{F}=\frac{1}{2}mv_{F}^{2}\equiv k_{B}T_{F}
$ where $T_{F}$\ is the Fermi temperature. The Fermi energy $E_{F}$ equals $%
\pi \hbar ^{2}n/m$ in 2D and $(\hbar ^{2}/2m)(3\pi ^{2}n)^{2/3}$ in 3D, with 
$n$ the total number-density of charge-carrier electrons, while $E_{f}$ is
the same with $n$ replaced by, say, $n_{f}$. The quantities $E_{f}$ and $%
E_{F}$ coincide {\it only }when perfect 2e/2h-CP symmetry holds, i.e.
when $n=n_{f}$.

The grand potential $\Omega $\ for the full $H=H_{0}+H_{int}$ is then
constructed via%
$$
\Omega (T,L^{d},\mu ,N_{0},M_{0}) \ =-k_{B}T\ln \left[ {\rm Tr}%
e^{-\beta (H-\mu \hat{N})}\right]   
\eqno{(5)}
$$%
where ``Tr'' stands for ``trace.'' Following the Bogoliubov recipe 
[39], one sets $b_{\bf{0}}^{+},$ $b_{\bf{0}}$ equal to $\sqrt{%
N_{0}}$ and $c_{\bf{0}}^{+}$, $c_{\bf{0}}$ equal to $\sqrt{M_{0}}$
in (1), where $N_{0}$ is the $T$-dependent number of zero-CMM
2e-CPs and $M_{0}$ the same for 2h-CPs. This allows {\it exact }%
diagonalization, through a Bogoliubov transformation, giving [40] 
$$
\eqalignno{
 \frac{\Omega}{L^{d}} = \int_{0}^{\infty
}d\epsilon N(\epsilon )[\epsilon -\mu -E(\epsilon )]-2k_{B}T\int_{0}^{\infty
}d\epsilon N(\epsilon )\ln \{1+\exp [-\beta E(\epsilon )]\} \qquad \qquad \cr
+[E_{+}(0)-2\mu ]n_{0}+k_{B}T\int_{0^{+}}^{\infty }d\varepsilon
M(\varepsilon )\ln \{1-\exp [-\beta \{E_{+}(0)+\varepsilon -2\mu \}]\} 
 \qquad \qquad \cr
\quad \quad +[2\mu -E_{-}(0)]m_{0}+k_{B}T\int_{0^{+}}^{\infty }d\varepsilon
M(\varepsilon )\ln \{1-\exp [-\beta \{{2\mu -E}_{-}(0)+\varepsilon \}]\} \quad \quad
(6)
}
$$
where $N(\epsilon )$ and $M(\varepsilon )$ are respectively the electronic
and bosonic density of states, $E(\epsilon )=\sqrt{(\epsilon -\mu
)^{2}+\Delta ^{2}(\epsilon )}$ where $\Delta (\epsilon )\equiv \sqrt{n_{0}}%
f_{+}(\epsilon )+\sqrt{m_{0}}f_{-}(\epsilon )$, with $n_{0}(T)\equiv
N_{0}(T)/L^{d}$ and $m_{0}(T)\equiv M_{0}(T)/L^{d}$ being the 2e-CP and
2h-CP number densities, respectively, of BE-condensed bosons. Minimizing (6)
 with respect to $N_{0}$ and $M_{0}$, while simultaneously
fixing the total number $N$ of electrons by introducing the electron
chemical potential $\mu $, namely 
$$
\frac{\partial \Omega }{\partial N_{0}} \ = \ 0,\, {\ \ \
\ \ \ }\frac{\partial \Omega }{\partial M_{0}} =  \ 0, {
\ \ \ \ \ \ \ {\rm and} \ \ \ \ \ \ \ \ }\frac{\partial \Omega }{\partial \mu } = \ -N  
\eqno{(7)}
$$%
specifies an {\it equilibrium state} of the system with volume $L^{d}$
and temperature $T$. Here $N$ evidently includes both paired and unpaired CP
electrons. The diagonalization of the CBFM $H$ is {\it exact}, unlike
with the BCS $H$, so that the CBFM goes beyond mean-field theory.{\it \ }%
Some algebra then leads [40] to the three coupled integral Eqs.
(7)-(9) of Ref. [12]. Self-consistent (at worst, numerical) solution
of these {\it three coupled equations} then yields the three
thermodynamic variables of the CBFM%
$$
\ \ n_{0}(T,n,\mu ),\;\ \ \ m_{0}(T,n,\mu ),\ \ \ \ {\rm and}\ \ \ \ \mu(T,n).  
\eqno{(8)}
$$%
Fig.1 displays the
three BE condensed phases---labeled $s+$, $s-$ and $ss$---along with the
normal phase $n$, that emerge [13] from the CBFM.

\topinsert
\input psfig.sty
\centerline{\hskip-12mm\psfig{figure=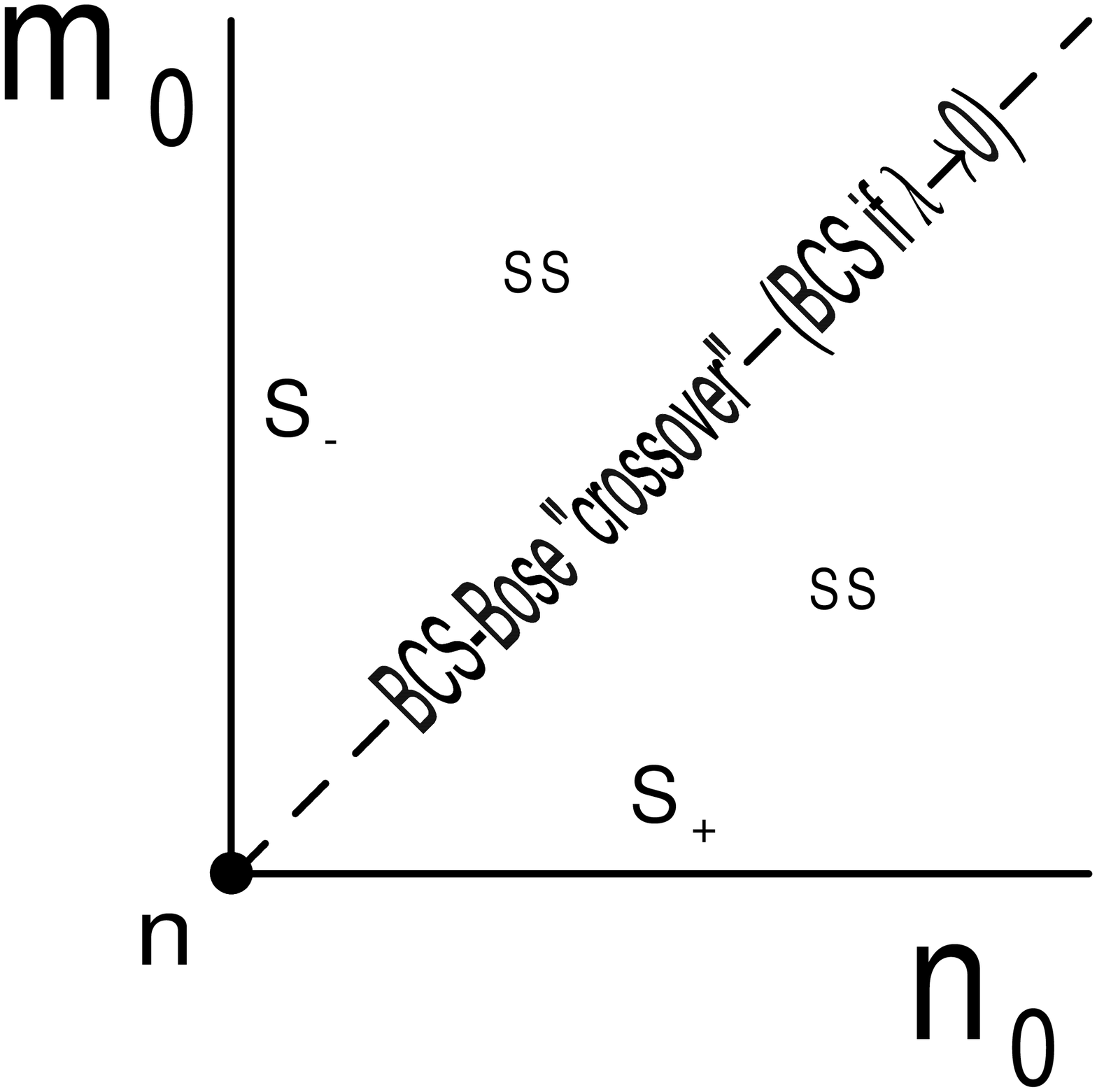,height=16truecm,width=12truecm,angle=0}}
\vskip -5.2truecm 
\noindent
{\bf Figure 1.} 
Illustration in the $n_{0}$-$m_{0}$ plane of three CBFM condensed phases
(the pure 2e-CP $s+$ and 2h-CP $s-$ BE condensate phases and a mixed phase $%
ss$) along with the normal (ternary BF non-Fermi-liquid) phase $n$.

\vskip 12truept
\endinsert

Vastly more general, the CBFM contains [1] the key equations of
all {\it five }distinct statistical theories as special cases; these
range from BCS to BEC theories, which are thereby unified by the CBFM.
Perfect 2e/2h CP symmetry signifies equal numbers of 2e- and 2h-CPs, more
specifically, $n_{B}(T)=m_{B}(T)$ {\it as well as }$n_{0}(T)=m_{0}(T).$
With (4) this implies that $E_{f}$ coincides with $\mu $,
and the CBFM then reduces to the gap and number equations [viz., (11) and (12)
 below]\ of the {\it BCS-Bose crossover
picture} with the Cooper/BCS model interaction---if its parameters $V$ and $%
\hbar \omega _{D}$ are identified with the BF interaction Hamiltonian $%
H_{int}$ parameters $f^{2}/2\delta \varepsilon $ and $\delta \varepsilon $,
respectively.\ The crossover picture for unknowns $\Delta (T)$ and $\mu (T)$
is now supplemented by the central relation 
$$
\Delta (T)=f\sqrt{n_{0}(T)}=f\sqrt{m_{0}(T)}.  
\eqno{(9)}
$$%
Both $\Delta (T)$\ \ and $n_{0}(T)$\ and $m_{0}(T)$\ are the familiar
``half-bell-shaped'' order-parameter curves. These are zero above a certain
critical temperature $T_{c}$, rising monotonically upon cooling (lowering $T$%
) to maximum values $\Delta (0),$\ $n_{0}(0)$\ and $m_{0}(0)$\ at\ $T=0.$
The energy gap $\Delta (T)$\ is the order parameter describing the
superconducting (or superfluid) condensed state,\ while $n_{0}(T)$\ and\ $%
m_{0}(T)$\ are the BEC order parameters depicting the macroscopic occupation
that arises below $T_{c}$ in a BE condensate. This $\Delta (T)$\ is
precisely the BCS energy gap if the boson-fermion coupling $f$ is made to
correspond to $\sqrt{2V\hbar \omega _{D}}$. Note that the BCS and BE $T_{c}$%
s are the same. Writing (9) for $T=0$, and dividing this
into (9) gives the much simpler $f$-independent relation
involving order parameters {\it normalized} in the interaval [$0,1$]%
$$
\Delta (T)/\Delta (0)= \sqrt{n_{0}(T)/n_{0}(0)}\ = \sqrt{m_{0}(T)/m_{0}(0)} \quad \smash
{\mathop{\relbar\joinrel\longrightarrow}\limits_{T \to 0}} \quad 1.  
\eqno{(10)}
$$%
The first equality, apparently first obtained in Ref. [9], simply
relates the two heretofore unrelated ``half-bell-shaped'' order parameters
of the BCS and the BEC theories.\ The second equality [12, 13]
implies that a BCS condensate is precisely a BE condensate of equal
numbers of 2e- and 2h-CPs. Since (10) is {\it independent }%
of the particular two-fermion dynamics of the problem, it can be expected to
hold for either SCs and SFs.

\vskip 28 truept

\centerline{\bf 3. BCS-BOSE CROSSOVER THEORY}
\vskip 12 truept

The crossover theory (defined by two simultaneous equations, the gap and
number equations) was introduced by many authors beginning in 1967 with
Friedel and co-authors [41]; for a review see Ref. [42].
The critical temperature $T_{c}$ is defined by $\Delta (T_{c})=0$, and is to
be determined self-consistently with $\mu (T_{c})$. The two equations to be
solved, in 2D for the Cooper/BCS model interaction, are [43] 
$$
1=\lambda \int_{0}^{\frac{\hbar \omega _{D}}{2k_{B}T_{c}}}dx\frac{\tanh x}{x}\ \ 
( {\rm if} \, \, \mu >\hbar \omega _{D} ); \quad  1 = \lambda \int_{\frac{-\mu
(T_{c})}{2k_{B}T_{c}}}^{\frac{\hbar \omega _{D}}{2k_{B}T_{c}}}dx\frac{\tanh x}{2x},\ \ 
({\rm if} \, \, \mu <\hbar \omega_{D})  
\eqno{(11)}
$$
$$
\int_{0}^{\infty }\frac{d\epsilon }{\exp [\epsilon -\mu
(T_{c})/2k_{B}T_{c}]+1}=1.  
\eqno{(12)}
$$%
The last integral can be done analytically, and leaves 
$$
\mu (T_{c})=k_{B}T_{c}\ln (e^{E_{F}/k_{B}T_{c}}-1).  \eqno{(13)}
$$%
The $\mu (T_{c})$ is then eliminated (numerically) from (11) to
give $T_{c}$ as a function of $\lambda $. Using $\hbar \omega _{D}/E_{F}=0.05
$ as a typical value for cuprates, increasing $\lambda $ makes $\mu (T_{c})$
decrease from its weak-coupling (where $T_{c}\rightarrow 0$) value of $E_{F}$
down to $\hbar \omega _{D}$ when $\lambda \simeq 56$, an unphysically large
value. Fig. 2 displays $T_{c}$ (in units of $T_{F}$) as function of $\lambda 
$. 

\topinsert
\input psfig.sty
\centerline{\psfig{figure=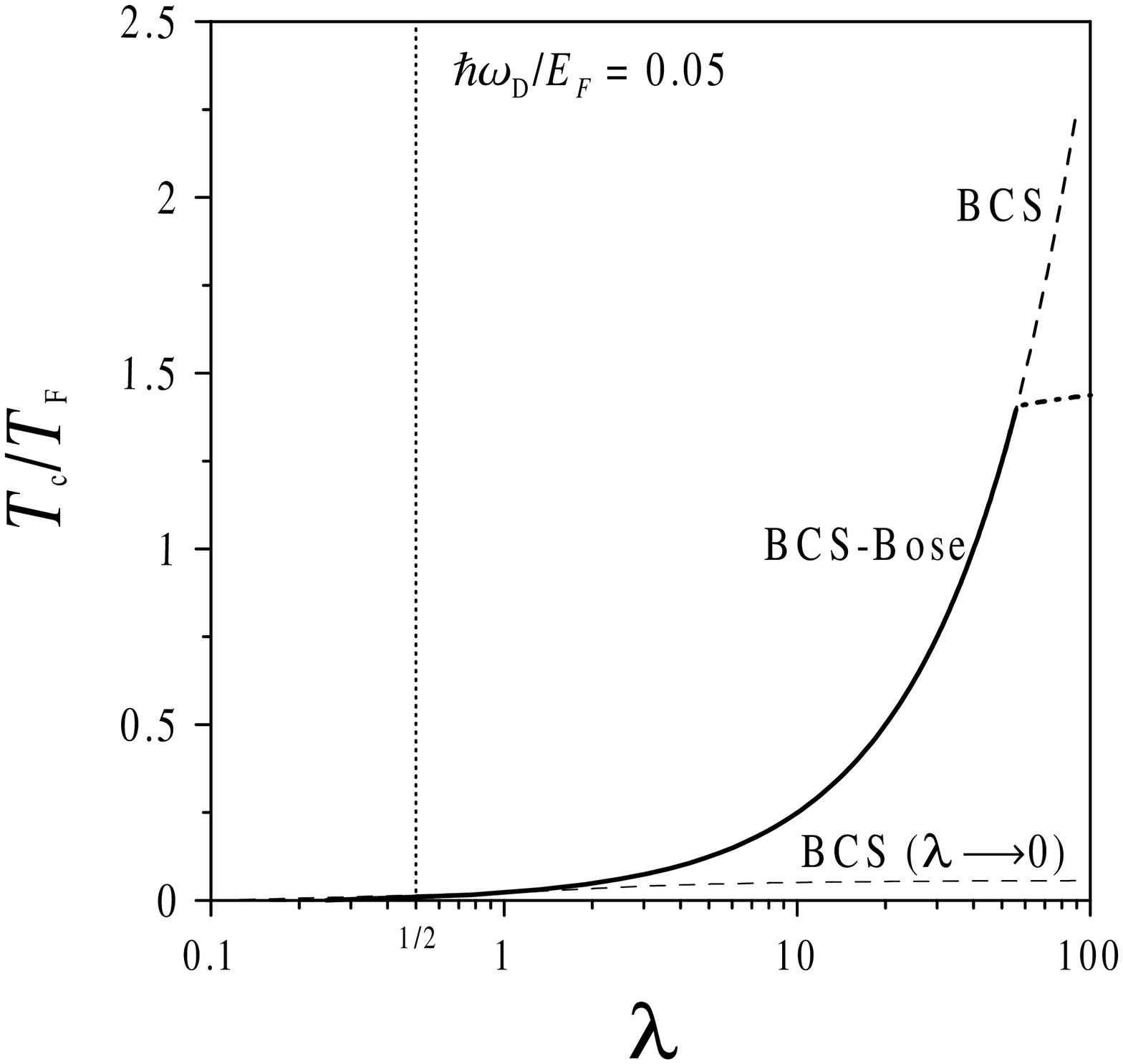,width=14truecm,height=12truecm}}
\vskip -1.4truecm
\noindent
{\bf Figure 2.}
Critical SC temperatures $T_{c}$ in units of $T_{F}$ for the BCS-Bose
crossover theory (full curve), the BCS value from the exact implicit equation (Ref. [21], p. 447) $1$ $=\lambda \int_{0}^{\hbar
\omega _{D}/2k_{B}T_{c}}dxx^{-1}\tanh x$ (upper dashed curve) and its
weak-coupling solution $T_{c}\simeq 1.134\hbar \omega _{D}\exp (-1/\lambda )$
(lower dashed curve). The dot-dashed ``appendage'' signals a breakdown in
the BCS/Cooper interaction model when $\mu (T_{c})$ turns negative, as the
Fermi surface at $\mu $ then washes out. The value of $\lambda =1/2$\ marked
is the maximum possible value allowed [45] for this interaction
model just short of lattice instability.
\vskip 12truept
\endinsert

\vskip 28 truept

\centerline{\bf 4.  GAP EQUATION}
\vskip 12 truept

Curiously, the standard procedure in all SC and SF theories of many-fermions
is to {\it ignore }2h-CPs altogether. Indeed, the BCS gap equation for
all $T$\ can be derived without them. Neglecting in (6) all terms
containing $m_{0}(T)$, $E_{-}(0)$ and $f_{-}(\epsilon )$ leaves an $\Omega
(T,L^{d},\mu ,N_{0})$ defining an {\it incomplete }BFM.\ Minimizing it
over $N_{0}$\ (for fixed total electron number $N$) requires that $\partial
\Omega /\partial N_{0}=0$ or $\partial \Omega /\partial n_{0}=0$, which becomes%
$$
\int_{0}^{\infty }d\epsilon N(\epsilon )\left[ -1+\frac{2\exp \{-\beta
E(\epsilon )\}}{1+\exp \{-\beta E(\epsilon )\}}\right] \frac{dE(\epsilon )}{%
dn_{0}}+\left[ E_{+}(0)-2\mu \right] =0
$$
or%
$$
2\left[ E_{+}(0)-2\mu \right] =f^{2}\int_{E_{f}}^{E_{f}+\delta \varepsilon
}d\epsilon N(\epsilon )\frac{1}{E(\epsilon )}\tanh 
{\frac12}%
\beta E(\epsilon ).
\eqno{(14)}
$$%
Using (4) yields precisely the BCS gap equation for all $T$, namely
$$
1=\lambda \int_{0}^{\hbar \omega _{D}}d\xi \frac{1}{\sqrt{\xi ^{2}+\Delta
^{2}(T)}}\tanh 
{\frac12}%
\beta \sqrt{\xi ^{2}+\Delta ^{2}(T)}  
\eqno{(15)}
$$%
where $\xi \equiv \epsilon -\mu $, provided one picks $E_{f}=\mu $, since $%
\lambda \equiv f^{2}N(0)/2\delta \varepsilon $ while $\delta \varepsilon
=\hbar \omega _{D}$.
The companion number equation follows from
the last equation of (7) and will thus be
$$
n=n_{f}(T)+2n_{B}(T)
\eqno{(16)}
$$
where $n_{f}(T)$ is the number density of unpaired electrons%
$$
n_{f}(T)\equiv \int\limits_{0}^{\;\infty }d\epsilon N(\epsilon )[1-\frac{%
\epsilon -\mu }{E(\epsilon )}\tanh {\frac12}
\beta E(\epsilon )]
\eqno{(17)}
$$
while the number density of composite bosons, both with $K=0$ and with $K>0$, is
$$
n_{B}(T)\equiv n_{0}(T)+n_{B+}(T); \qquad n_{B+}(T)\equiv
\int\limits_{0+}^{\infty }d\varepsilon M(\varepsilon )\frac{1}{e^{\beta
\lbrack E_{+}(0) - 2\mu + \varepsilon]}-1}.
\eqno{(18)}
$$
Note that the number equation $n=n_{f}(T)+2n_{B}(T)$\ {\it{differs}}\ from
(12) of the previous section which follows from $n=n_{f}(T)$
only.\ 

Similarly, ignoring 2e-CPs and keeping only 2h-CPs leads to $\Omega
(T,L^{d},\mu ,M_{0})$ which to minimize over $M_{0}$ requires that 
$\partial \Omega /\partial M_{0}=\partial \Omega /\partial m_{0}=0$ and,
since $E(\xi )\equiv E(-\xi )$. This {\it{again}} leads to (15)
but now with the companion number equation 
$$
m=n_{f}(T)-2m_{B}(T)
\eqno{(19)}
$$
with the same previous $n_{f}(T)$ and where
$$
m_{B}(T)\equiv m_{0}(T) + m_{B+}(T); \qquad m_{B+}(T)\equiv
\int\limits_{0+}^{\infty }d\varepsilon M(\varepsilon )\frac{1}{e^{\beta
\lbrack {2\mu } - E_{-}(0) + \varepsilon ]}-1}.
\eqno{(20)}
$$
However, ignoring either 2e- or 2h-CPs does {\it{not}} give the entire BCS
ground-state energy, as we now show.

\vskip 28 truept

\centerline{\bf 5.  CONDENSATION ENERGY}
\vskip 12 truept

The $T=0$ condensation energy per unit volume according to the CBFM (i.e.,
with {\it both }2e- and 2h-CPs)\ is%
$$
\frac{E_{s}-E_{n}}{L^{d}}=\frac{\Omega _{s}(T=0)-\Omega _{n}(T=0)}{L^{d}}
\eqno{(21)}
$$%
since for any $T$\ the Helmholtz free energy $F=\Omega +\mu N=E-TS$, with $S$
the entropy, and $\mu $ is the same for either superconducting $s$ or normal 
$n$ phases with internal energies $E_{s}$ and $E_{n}$, respectively. In the 
{\it normal} phase $n_{0}=0$, $m_{0}=0$ so that $\Delta (T)=0$ for all $%
T\geq 0$, and (6) reduces to 
$$
\frac{\Omega _{n}(T=0)}{L^{d}}=\int_{0}^{\infty }d\epsilon N(\epsilon
)[\epsilon -\mu -|\epsilon -\mu |]=2\int_{0}^{\mu }d\epsilon N(\epsilon
)[\epsilon -\mu ]=2\int_{-\mu }^{0}d\xi N(\xi )\xi .  \eqno{(22)}
$$
For the superconducting phase at $T=0,$ and when $n_{0}(T)=m_{0}(T)$ and $%
n_{B}(T)=m_{B}(T)$ hold, one deduces from (4) and (6) 
that $\mu =E_{f}$. Letting $\Delta (T=0)$ $\equiv \Delta $ in (6) 
and putting $\delta \varepsilon \equiv \hbar \omega _{D}$ while
using (4) gives for the superconducting phase%
$$
\eqalignno{
\frac{\Omega _{s}(T=0)}{L^{d}} = \, \, \, & 2\hbar \omega _{D}n_{0}(0)+\int_{-\mu
}^{\infty }d\xi N(\xi )\left( \xi -\sqrt{\xi ^{2}+\Delta ^{2}}\right)  
  \cr
\qquad \qquad  = \, \, \, & 2\hbar \omega _{D}n_{0}(0)+2\int_{-\mu }^{-\hbar \omega _{D}}d\xi N(\xi
)\xi -2\int_{0}^{\hbar \omega _{D}}d\xi N(\xi )\sqrt{\xi ^{2}+\Delta ^{2}}.
  \qquad  \, \, \, \, 
(23)
}
$$

Subtracting (22) from (23) and putting $N(\xi )\cong
N(0)$, the density of electronic states at the Fermi surface, one is left
with%
$$
\frac{E_{s}-E_{n}}{L^{d}} \ =\ 2\hbar \omega
_{D}n_{0}(0)+2N(0)\int_{0}^{\hbar \omega _{D}}d\xi \left( \xi -\sqrt{\xi
^{2}+\Delta ^{2}}\right) \quad \quad \hbox{(CBFM)}. 
 \eqno{(24)}
$$%
Employing Eq. (2), p. 158 of Ref. [44] the integral becomes 
$$
\eqalignno{
\qquad \frac{(\hbar \omega _{D})^{2}}{2}-\frac{1}{2}\hbar \omega _{D}\sqrt{(\hbar
\omega _{D})^{2}+\Delta ^{2}}+\frac{1}{2}\Delta ^{2}\ln \frac{\Delta }{\hbar
\omega _{D}+\sqrt{(\hbar \omega _{D})^{2}+\Delta ^{2}}} &  \cr
\smash {\mathop{\relbar\joinrel\longrightarrow}\limits_{\Delta \to 0}}%
\quad \frac{1}{2}\Delta ^{2}\ln \left( \frac{\Delta }{2\hbar \omega
_{D}}\right) -\frac{1}{4}\Delta ^{2}-\frac{1}{16}\frac{\Delta ^{4}}{(\hbar
\omega _{D})^{2}}+O\left( \frac{\Delta ^{6}}{\left[ \hbar \omega _{D}\right]
^{4}}\right). & \qquad \qquad (25)
}
$$
Using (9) for $T=0$ and weak coupling $f\rightarrow 0$
implies that $\Delta =f\sqrt{n_{0}(0)}=f\sqrt{m_{0}(0)}\rightarrow 0$ so
that (24) yields the expansion%
$$
\eqalign{
& \frac{E_{s}-E_{n}}{L^{d}} \quad  \smash
{\mathop{\relbar\joinrel\longrightarrow}\limits_{\Delta \to 0}} \quad
2\hbar \omega _{D}n_{0}(0)   \cr
& +2N(0)\left[ \frac{1}{2}\Delta ^{2}\ln \left( \frac{\Delta }{2\hbar \omega
_{D}}\right) -\frac{1}{4}\Delta ^{2}-\frac{1}{16}\frac{\Delta ^{4}}{(\hbar
\omega _{D})^{2}}+O\left( \frac{\Delta ^{6}}{\left[ \hbar \omega _{D}\right]
^{4}}\right) \right] \quad \hbox{(CBFM)}. 
\quad {(26)}
}
$$
Given that for small $\lambda $ 
$$
\Delta =\frac{\hbar \omega _{D}}{\sinh (1/\lambda )}\quad \smash
{\mathop{\relbar\joinrel\longrightarrow}\limits_{\lambda \to 0}} \quad 2\hbar \omega _{D}\exp (-1/\lambda )
\eqno{(27)}
$$%
the log term in (26) is just%
$$
\ln \left( \frac{\Delta }{2\hbar \omega _{D}}\right) =-\frac{2\hbar \omega
_{D}}{f^{2}N(0)}
\eqno{(28)}
$$%
since 
$$
\lambda \equiv VN(0)=\frac{f^{2}N(0)}{2\hbar \omega _{D}}
\eqno{(29)}
$$%
so that (26) finally simplifies to%
$$
\frac{E_{s}-E_{n}}{L^{d}} \quad \smash
{\mathop{\relbar\joinrel\longrightarrow}\limits_{\Delta \to 0}} \quad 
- \frac{1}{2}N(0)\Delta ^{2}\left[ 1+\frac{1}{4}\left( \frac{\Delta }{\hbar
\omega _{D}}\right) ^{2}+O\left( \frac{\Delta }{\hbar \omega _{D}}\right)
^{4}\right] \quad \quad \hbox{(CBFM)}.  
\eqno{(30)}
$$

By contrast, the original BCS expression from Eq. (2.42) of Ref. [6]
 is 
$$
\frac{E_{s}-E_{n}}{L^{d}}=N(0)(\hbar \omega _{D})^{2}\left[ 1-\sqrt{1+\left(
\Delta /\hbar \omega _{D}\right) ^{2}}\right] \quad \quad \hbox{(BCS)}
\eqno{(31)}
$$%
which on expansion leaves%
$$
\frac{E_{s}-E_{n}}{L^{d}} \quad \smash
{\mathop{\relbar\joinrel\longrightarrow}\limits_{\lambda \to 0}} \quad -\frac{1}{2}%
N(0)\Delta ^{2}\left[ 1-\frac{1}{4}\left( \frac{\Delta }{\hbar \omega _{D}}%
\right) ^{2}+O\left( \frac{\Delta }{\hbar \omega _{D}}\right) ^{4}\right] 
\quad \quad \hbox{(BCS)}.  
\eqno{(32)}
$$%
Thus, the CBFM condensation energy (30), and consequently its
ground-state energy, is {\it lower} (or larger in magnitude) than the BCS
result (32). Therefore, the CBFM satisfies a prime
expectation of any theory that improves upon BCS, which being based on a
trial wave function gives a ground-state energy which is a rigorous upper
bound to the exact energy associated with the BCS Hamiltonian ground state.
Consequently, there is no {\it a priori }reason why the CBFM is limited
to weak coupling, at least for all $\lambda \leq 1/2$ [45].

What happens on ignoring {\it either }2e- or 2h-CPs, as seems to be
common practice in theories of SCs and SFs? Starting from (6) for $%
T=0$, and following a similar procedure to arrive at (23) but 
{\it without} 2h-CPs such that $f_{-}=0$, $m_{0}(0)=0$ and $n_{0}(0)=\Delta ^{2}/f^{2}$, one gets 
$$
\left[ \frac{\Omega _{s}(T=0)}{L^{d}}\right] _{+}=\ \hbar \omega
_{D}n_{0}(0)+2\int_{-\mu }^{0}d\xi N(\xi )\xi +N(0)\int_{0}^{\hbar \omega
_{D}}d\xi \left( \xi -\sqrt{\xi ^{2}+\Delta ^{2}}\right) .  
\eqno{(33)}
$$%
Subtracting (22) from (33) gives%
$$
\left[ \frac{E_{s}-E_{n}}{L^{d}}\right] _{+}=\hbar \omega
_{D}n_{0}(0)+N(0)\int_{0}^{\hbar \omega _{D}}d\xi \left( \xi -\sqrt{\xi
^{2}+\Delta ^{2}}\right)   
\eqno{(34)}
$$%
which is just {\it one half }the full CBFM\ result (24).
Furthermore, if $[(E_{s}-E_{n})/L^{d}]_{-}$ is the contribution from 2h-CPs
alone, assuming now that $f_{+}=0$ and $n_{0}(0)=0$\ we eventually arrive at
precisely rhs of (34) but with $m_{0}(0)=\Delta ^{2}/f^{2}$\ in
place of $n_{0}(0)$. Hence%
$$
\left[ \frac{E_{s}-E_{n}}{L^{d}}\right] _{+} = \left[ \frac{E_{s}-E_{n}}{%
L^{d}}\right] _{-}\smash
{\mathop{\relbar\joinrel\longrightarrow}\limits_{\lambda \to 0}}  
 -\frac{1}{4}N(0)\Delta ^{2}\left[ 1+\frac{1}{4}\left( \frac{\Delta }{\hbar
\omega _{D}}\right) ^{2}+O\left( \frac{\Delta }{\hbar \omega _{D}}\right)
^{4}\right]
\hfill (35)
$$ 
which again is just one half the full CBFM condensation energy (30) 
that in leading order in $\Delta $ was found to be the full
BCS condensation energy.

Including both 2e- and 2h-CPs gave similarly striking conclusions on
generalizing [22, 23] the ordinary [18] CP
problem from unrealistic infinite-lifetime pairs to the physically expected
finite-lifetime ones.\ 

\vskip 28 truept

\centerline{\bf 6.  CONCLUSIONS}
\vskip 12 truept

The recent ``complete boson-fermion model'' (CBFM) contains as a special
case the BCS-Bose crossover theory which, at least for the Cooper/BCS model
interaction, predicts virtually the same $T_{c}$s to well beyond physically
unreasonable values of coupling than the allegedly less general BCS theory
where the number equation is replaced by the assumption that $\mu =E_{F}$ . 

The CBFM reveals that, while the BCS gap equation for all
temperatures follows rigorously without either electron or hole pairs, the
resulting $T=0$ condensation energy is only one half the entire BCS value.
In view of this, if BEC is at all relevant in SCs and SFs taken as
many-fermion systems where pairing into bosons undoubtedly occurs, two-hole
CPs cannot and must not be ignored. 

\vskip 28 truept

\centerline{\bf ACKNOWLEDGMENTS}
\vskip 12 truept
We thank M. Fortes, J. Javanainen, O. Rojo and
V.V. Tolmachev for extensive discussions and acknowledge UNAM-DGAPA-PAPIIT
(Mexico) grant IN106401, and CONACyT (Mexico) grants 41302 and 43234-F, for partial
support. MdeLl is grateful for travel support through a grant to Southern
Illinois University at Carbondale from the U.S. Army Research Office.

\vskip 28 truept

\centerline{\bf REFERENCES}
\vskip 12 truept


\item{[1]} M. de Llano, in {\it Frontiers in Superconductivity
Research }(ed. B.P. Martins) (Nova Science Publishers, NY, 2004). Available
in cond-mat/0405071.

\item{[2]} J.M. Blatt, {\it Theory of Superconductivity} (Academic,
New York, 1964).

\item{[3]} M.R. Schafroth, Phys. Rev. {\bf 96}, 1442 (1954).

\item{[4]} M.R. Schafroth, {\it et al.}, Helv. Phys. Acta {\bf 30},
93\ (1957).

\item{[5]} M.R. Schafroth, Sol. State Phys. {\bf 10}, 293 (1960).

\item{[6]} J. Bardeen, L.N. Cooper \& J.R. Schrieffer, Phys. Rev. {\bf 108}, 1175\ (1957).

\item{[7]} N.N. Bogoliubov, JETP {\bf 34}, 41 (1958).

\item{[8]} N.N. Bogoliubov, V.V. Tolmachev \& D.V. Shirkov, Fortschr.
Phys. {\bf 6}, 605 (1958); and also in {\it A New Method in the Theory
of Superconductivity} (Consultants Bureau, NY, 1959).

\item{[9]} J. Ranninger \& S. Robaszkiewicz, Physica B {\bf 135}, 468
(1985).

\item{[10]} R. Friedberg \& T.D. Lee, Phys. Rev. B{\bf \ 40}, 6745
(1989).

\item{[11]} R. Friedberg, {\it et al.}, Phys. Lett. A {\bf 152}, 417
and 423 (1991).

\item{[12]} V.V. Tolmachev, Phys. Lett. A {\bf 266}, 400 (2000).

\item{[13]} M. de Llano \& V.V. Tolmachev, Physica A {\bf 317}, 546
(2003){\it .}

\item{[14]} J. Batle, {\it et al.}, Cond. Matter Theories {\bf 18}%
, 111 (2003){\it .} Cond-mat/0211456.

\item{[15]} M. Casas, {\it et al.}, Phys. Lett. A {\bf 245}, 5
(1998).

\item{[16]} M. Casas, {\it et al.}, Physica A {\bf 295}, 146 (2001).

\item{[17]} M. Casas, {\it et al.}, Sol. State Comm. {\bf 123}, 101
(2002).

\item{[18]} L.N. Cooper, Phys. Rev. {\bf 104}, 1189 (1956).

\item{[19]} S.K. Adhikari, {\it et al.}, Phys. Rev. B {\bf 62},
8671 (2000).

\item{[20]} S.K. Adhikari, {\it et al.}, Physica C {\bf 351},
341 (2001).

\item{[21]} A.L. Fetter \& J.D. Walecka, {\it Quantum Theory of
Many-Particle Systems} (McGraw-Hill, New York, 1971).

\item{[22]} M. Fortes, {\it et al.}, Physica C {\bf 364-365},
95 (2001).

\item{[23]} V.C. Aguilera-Navarro, {\it et al., }Sol. St. Comm. 
{\bf 129}, 577 (2004).

\item{[24]} N.N. Bogoliubov, N. Cim.{\bf \ 7}, 794 (1958).

\item{[25]} D. Vollkardt \& P. W\"{o}lfle, {\it The Superfluid Phases
of Helium 3} (Taylor \& Francis, London, 1990).

\item{[26]} E.R. Dobbs, {\it Helium Three} (Oxford University Press,
Oxford, UK, 2000).

\item{[27]} M.J. Holland, B. DeMarco \& D.S. Jin, Phys. Rev. A {\bf %
61}, 053610 (2000).

\item{[28]} K.M. O'Hara, S.L. Hammer, M.E. Ghem, S.R. Granade \& J.E.
Thomas, Science {\bf 298,} 2179 (2002).

\item{[29]} K.E. Strecker, G.B. Partridge \& R.G. Hulet, Phys. Rev. Lett. 
{\bf 91}, 080406 (2003).

\item{[30]} M. Holland, S.J.J.M.F. Kokkelmans, M.L. Chiofalo \& R.
Walser, Phys. Rev. Lett. {\bf 87}, 120406 (2001).

\item{[31]} E. Timmermans, K. Furuya, P.W. Milonni \& A.K. Kerman,
Phys. Lett. A {\bf 285}, 228 (2001).

\item{[32]} M.L. Chiofalo, S.J.J.M.F. Kokkelmans, J.N. Milstein \&
M.J. Holland, Phys. Rev. Lett. {\bf 88}, 090402 (2002).

\item{[33]} Y. Ohashi \& A. Griffin, Phys. Rev. Lett. {\bf 89},
130402 (2002).

\item{[34]} L. Pitaevskii \& S. Stringari, Science {\bf 298},
2144 (2002).

\item{[35]} M. Greiner, C.A. Regal \& D.S. Jin, Nature {\bf 426}, 537
(2003).

\item{[36]} C.A. Regal, M. Greiner \& D.S. Jin, Phys. Rev. Lett. {\bf %
92}, 040403 (2004).

\item{[37]} M.W. Zwierlein, C. A. Stan, C. H. Schunck, S.M. F.
Raupach, S. Gupta, Z. Hadzibabic \& W. Ketterle, Phys. Rev. Lett. {\bf 91}%
, 250401 (2003).

\item{[38]} S. Jochim, M. Bartenstein, A. Altmeyer, G. Hendl, S. Riedl,
C. Chin, J. Hecker Denschlag \& R. Grimm, Science {\bf 302}, 2101 (2003).

\item{[39]} N.N. Bogoliubov, J. Phys. (USSR) {\bf 11}, 23 (1947).

\item{[40]} M. Casas, M. de Llano \& V.V. Tolmachev, {\it to be
published.}

\item{[41]} J. Labb\'{e}, S. Barisic \& J. Friedel, Phys. Rev. Lett. 
{\bf 19}, 1039\ (1967).

\item{[42]} M. Randeria, in {\it Bose-Einstein Condensation}, ed.
A. Griffin {\it et al.} (Cambridge University, Cambridge, 1995) p. 355{.}

\item{[43]} F.J. Sevilla, PhD Thesis (UNAM, 2004) {\it %
unpublished.}

\item{[44]} A. Jeffrey, {\it Handbook of Mathematical Formulas and
Integrals} (Academic Press, Newcastle, UK, 1995).

\item{[45]} A.B. Migdal, JETP {\bf 7}, 996 (1958).

\end{document}